# Beyond the Pandemic:

# Transforming Software Development with the IJARS Model for Wellbeing and Resilience

Daniel Russo
Aalborg University
daniel.russo@cs.aau.dk

**Abstract** - The recent global workforce upheaval, primarily instigated by the COVID-19 pandemic, has significantly impacted the software development sector. As developers adapted to remote work environments, the industry confronted profound challenges relating to employee wellbeing and resilience. This article provides an in-depth exploration into the transformative lessons of this period, emphasizing the necessity of understanding and fostering software developers' wellbeing. Through extensive research, we introduce the Integrated Job Demands-Resources and Self-Determination Model (IJARS), offering a nuanced understanding of wellbeing and productivity during the pandemic. We also uncover the pivotal role of Agile values, the imperative of proactive mental health measures, and the importance of learning from disruptions to build a robust workforce. The insights emphasize the need to reshape workplaces, promoting work-life balance, leveraging hybrid work models, and preparing for potential future upheavals. This article provides guidance towards a resilient and adaptive future, transforming adversity into opportunity.

**Introduction: Navigating the New Normal in Software Development**

The past few years have witnessed a seismic shift in the global workforce, driven by disruptive events such as the COVID-19 pandemic, geopolitical strife, and an increasing focus on sustainability and climate change. These challenges have dramatically altered the landscape of work across all sectors, marking a paradigm shift from traditional work norms. However, the software development industry has felt the ripple effects acutely.

It is important to understand that the transition to remote work was not merely a change in the physical location of work; it was a fundamental disruption of work practices, team dynamics, and organizational structures. This shift has had profound implications for developers' wellbeing, mental health, and productivity, thereby making it imperative for organizations to reassess their strategies and practices.

Our research journey involved a comprehensive exploration of various aspects influencing the wellbeing and resilience of developers. We examined the work environment, mental health support systems, job expectations, personal circumstances, and the role of Agile values. We also introduced the Integrated Job Demands-Resources and Self-Determination Model (IJARS) as a framework to understand wellbeing and productivity.

This article distills key insights, underscoring the pivotal role of wellbeing and resilience in the software industry from our personal research experience. The lessons learned present opportunities to reshape workplaces, implement mental health measures, promote work-life balance, leverage hybrid work arrangements, and prepare for future disruptions.

As we move forward, the strategies and practices adopted in the software development industry have broader implications. By sharing our insights, we contribute to the discourse on the future of work, guiding sectors towards a resilient and adaptive future.

## Emphasizing Wellbeing for Productivity in Resilient Organizations

In the midst of a rapidly changing global landscape, the profound impact of a developer's wellbeing on their productivity has become glaringly evident. Our two-year longitudinal study (Russo et al., 2023a) painted a clear picture: when developers' wellbeing suffered during the pandemic, so did their productivity. On the flip side, those with high wellbeing levels consistently performed better, a finding that aligns with the results of a longitudinal study by Besker, Martini, and Bosch (2018) on developers' daily software development work.

This intricate balance between wellbeing and productivity cannot be overemphasized, especially in the context of software development. Mental health is not a mere supplementary concern—it stands at the heart of productivity and operational success (Bubonya et al., 2017). It is not just about bouncing back from challenges; it is about bouncing forward, absorbing shocks, and turning adversities into pathways for sustainable growth. As businesses grapple with multifaceted disruptions—from geopolitical tensions to talent attrition and the evident impacts of climate change—the need for organizational resilience becomes paramount.

Industry leaders, therefore, must acknowledge the central role mental health plays in achieving this resilience and driving productivity. While reactive measures to address mental health issues are a start, they are far from enough. The era calls for a proactive stance, where mental health in not just addressed but prioritized. This involves crafting a supportive work environment that champions mental wellbeing, flexibility, resource availability, and a culture of support and psychological safety (Edmondson, 1999).

Resilient organizations, do not merely respond to disruptions; they anticipate and adapt, turning challenges into opportunities (Yuan et al., 2022). This proactive resilience is evident in businesses that maintained strong organizational health during crises like the COVID-19 pandemic. By fostering knowledge sharing, performance reviews, and bottom-up innovation, they not only survived but thrived, outpacing competitors and setting themselves up for sustainable success.

# Hybrid Work Arrangements: The Future of Resilient Work-Life Balance

The future of work is in flux, and standing at its forefront is the enticing promise of hybrid work arrangements. Blending in-office collaboration with the autonomy of remote work offers a unique solution to modern challenges. It promises not only a better work-life balance but also a means to enhance developer wellbeing, subsequently boosting productivity and job satisfaction (Russo et al., 2023).

Our research (Šmite et al., 2023), coupled with recent industry insights, indicates that hybrid work models are more than just a trend; they are a transformative shift. These arrangements significantly amplify the work-life balance, driving not only the wellbeing of developers but also catalyzing their productivity and overall job satisfaction. Yet, the success of this model is not guaranteed—it hinges on thoughtful planning, transparent communication, and a robust support system.

Building resilience in this hybrid era demands a reimagining of traditional work structures. Organizations need to be Agile, ready to adapt swiftly to change. Teams should be self-sufficient, empowered with the autonomy to strategize and innovate. Leaders must be adaptable, guiding their teams through evolving landscapes, fostering new behaviors, and setting the foundation for both immediate responses and long-term resilience.

Furthermore, talent and culture become pivotal cornerstones. Resilient organizations attract top talent by offering adaptable environments. As the post-pandemic world grapples with unprecedented talent attrition, a hybrid model serves as both a solution and an attraction. These work models, backed by organizational resilience, become magnets for talent, drawing in those who value flexibility, growth, and wellbeing (Stander & Zyl, 2023).

Moreover, the transition to hybrid work models reflects a broader trend in organizational resilience. The multiple disruptions of recent years, from geopolitical events to global health crises, have underscored the importance of agility and adaptability. Resilient organizations do not just weather these storms; they harness them, leveraging challenges as catalysts for innovation and growth. By prioritizing leadership development and emphasizing adaptability, these organizations not only navigate disruptions but thrive amidst them (Stander & Zyl, 2023).

Theintertwined narratives of wellbeing, productivity, and hybrid work models are quite straightforward. But it is not just about adopting these models; it is about doing so with resilience at the core. It is about viewing disruptions not as setbacks but as opportunities, reshaping the workplace in ways that prioritize developer wellbeing, and building a future where challenges are the stepping stones to innovation and growth.

# A Holistic Approach to Employee Satisfaction in Resilient Organizations

Understanding the multifaceted elements that influence developers' wellbeing is paramount for organizations striving for resilience. Our research revealed that a multitude of factors contribute to a developer's wellbeing, extending far beyond the immediate impact of disruptive events such as the pandemic (Russo et al., 2023; Russo et al., 2021a). It became evident that to truly address and enhance developer wellbeing, organizations need to adopt a comprehensive, holistic approach that considers a wide spectrum of elements (Russo et al., 2023).

Employee satisfaction and wellbeing are intrinsically connected to various aspects of their work environment and personal life (Carver, Scheier, & Weintraub, 1989). As such, a one-dimensional approach that focuses solely on work-related factors would be inadequate. Instead, organizations should seek to understand and address this complex interplay of factors, which range from the physical work environment and job expectations to support systems and the personal circumstances of the employees.

The physical work environment plays a pivotal role in shaping employee satisfaction and productivity (Russo et al., 2023, Šmite et al., 2023). For developers, this could mean having a dedicated workspace, access to necessary tools and technologies, and an environment free from distractions. Resilient organizations recognize this and aim to create workspaces that foster creativity, collaboration, and productivity. This is particularly important in the context of hybrid work models, where the physical work environment could range from a traditional office to a home office setup.

Job expectations, too, significantly impact a developer's wellbeing and productivity. Clearly communicated, reasonable, and well-defined job expectations can help employees understand their roles, identify their goals, and measure their progress (Gittell, 2016). In contrast, unclear or unrealistic expectations can lead to job dissatisfaction, stress, and burnout. Resilient organizations understand this dynamic and prioritize clear communication about job expectations, roles, and responsibilities.

Support systems within an organization are another vital aspect of employee wellbeing. These systems can range from technical support to mental health resources and professional development opportunities (Gittell et al., 2016). Having robust support systems in place not only helps employees navigate their daily tasks but also empowers them to grow professionally and personally. Importantly, in resilient organizations, these support systems are not just present but are easily accessible and actively promoted.

The personal circumstances of employees also have a considerable impact on their wellbeing. These circumstances can include personal health, family responsibilities, financial stability, and more. Recognizing these factors and providing the necessary accommodations is a hallmark of resilient organizations (Gittell, 2016). Whether it is flexible work hours for parents, mental health support for those dealing with personal issues, or financial counseling services, these accommodations can make a significant difference in an employee's wellbeing and job satisfaction.

This holistic approach to employee wellbeing and satisfaction also aligns with the four key capabilities that resilient organizations focus on: building an Agile organization, fostering self-sufficient teams, promoting adaptable leaders, and investing in talent and culture. Each of these capabilities intersects with the various elements influencing employee wellbeing.

For instance, an Agile organization that can quickly adapt to changes offers employees a sense of stability and security, positively influencing their wellbeing. Similarly, self-sufficient teams empower employees, giving them a sense of ownership and control over their work, which can lead to increased job satisfaction. Adaptable leaders who guide and support their teams through change can help mitigate stress and uncertainty, enhancing employee wellbeing. Lastly, investing in talent and culture ensures that the work environment is supportive, inclusive, and growth-oriented, all of which contribute to employee satisfaction and wellbeing.

Indeed, cultivating resilience in organizations extends beyond merely navigating disruptions—it involves taking a holistic approach to employee wellbeing. It is about recognizing the various factors that influence employee satisfaction and wellbeing, and proactively addressing them. By adopting this holistic approach, organizations can not only enhance employee satisfaction and wellbeing but also boost productivity, foster innovation, and ultimately, build a more resilient organization.

## The Role of Agile values in Fostering Resilience

Agile values have emerged as effective approaches for navigating the challenges posed by remote work, a finding that is consistent across our research (Cucolaș & Russo, 2023). Agile methodologies, characterized by their focus on iterative development, collaboration, and adaptability, appear to align well with the demands of remote work. These principles are intrinsically geared towards resilience, allowing teams to rapidly respond to changes, ensuring continuity and efficiency even in the face of disruption.

Our research showed that Scrum, a popular Agile framework, has proven especially beneficial in remote work settings (Cucolaș & Russo, 2023). Its structured yet flexible approach provides a solid foundation for teams to collaborate, communicate, and improve overall performance, irrespective of their physical location. The way Scrum breaks work down into short, manageable sprints enables teams to adapt quickly to changes, receive regular feedback, and maintain a consistent work pace. This approach ensures that work is continuously progressing, and adjustments can be made quickly as new information or challenges arise.

In Scrum, the defined roles and rituals like the Scrum Master, Product Owner, daily stand-ups, sprint planning, and sprint retrospectives, facilitate better coordination, transparency, and learning among team members. The Scrum Master ensures that the team adheres to Scrum principles, removes impediments and facilitates team interactions. The Product Owner is responsible for maximizing the value of the product, managing the product backlog and ensuring that the team is working on the highest value features. These roles and rituals promote transparency and foster a culture of continuous learning and improvement, key characteristics of resilient organizations.

Building on this, resilient organizations often expand the application of Agile principles beyond their development teams. They incorporate the Agile mindset into their overall organizational culture. This means promoting values like adaptability, collaboration, customer-centricity, and continuous improvement throughout the organization. By doing so, they can foster an environment that is well-equipped to navigate changes and disruptions, making the organization as a whole more resilient.

An Agile organization can quickly pivot in response to changes, whether they are in customer preferences, market conditions, or internal processes (Stander & Zyl, 2023). Agile teams regularly reflect on their performance and seek ways to improve, ensuring that they are always learning and improving (Verwijs & Russo, 2023). This continuous learning cycle is a critical aspect of resilience. It allows organizations to not just bounce back from disruptions, but to learn from them and bounce forward, coming out stronger and better equipped for future challenges.

Agility also promotes a high level of customer-centricity. Agile teams work closely with customers, regularly gathering feedback and incorporating it into their products. This close relationship with the customer allows Agile organizations to better understand and respond to their customers' needs, leading to higher customer satisfaction and loyalty. In times of disruption, this customer-centric approach can help organizations maintain a strong customer relationship, ensuring business continuity and resilience.

Moreover, Agile fosters a culture of empowerment and autonomy. Agile teams are self-organizing, meaning they have the freedom to decide how they will work to achieve their goals. This autonomy can lead to higher job satisfaction, which in turn can boost productivity and resilience. In times of disruption, empowered teams can quickly adapt and find innovative solutions, without being hampered by rigid hierarchical structures.

The adoption of Agile values can significantly contribute to an organization's resilience. Through their focus on adaptability, continuous learning, customer-centricity, and team empowerment, Agile values provide a robust methodology for organizations to navigate disruptions. As we move into an increasingly unpredictable and fast-paced world, the role of Agile values in fostering resilience will only become more critical. By embracing Agile, organizations can equip themselves with the tools and mindset needed to turn challenges into opportunities and drive sustainable, inclusive growth.

# Embracing Disruptions as Learning Opportunities

The global COVID-19 pandemic, while devastating in many aspects, has also served as a catalyst for significant learning. Among the key lessons it has imparted, the importance of resilience in the face of disruptions stands out prominently (Russo et al., 2023a). The pandemic has had a substantial impact on software engineering practices, leading to considerable repercussions for the wellbeing and productivity of professionals in the sector.

These challenges have laid bare the need for robust strategies to safeguard and enhance wellbeing and productivity during times of crisis (Russo et al., 2023a). It has become evident that organizations need contingency plans that ensure not only the continuity of operations but also the support and wellbeing of their employees during such disruptions. Such plans may incorporate strategies like flexible work policies, virtual collaboration tools, mental health support, and clear communication channels (Russo et al., 2023).

Resilient organizations do not merely bounce back from adversities; they bounce forward. They absorb the shocks and transform them into opportunities to capture sustainable, inclusive growth (Gittell, 2016). When challenges emerge, leaders and teams in resilient organizations swiftly assess the situation, reorient themselves, double down on what is working, and discard what is not. The COVID-19 pandemic is a prime example of such an adversity that, despite its significant disruptions, has also served as a catalyst for organizations to reassess and reinvent their operational paradigms.

A crisis often brings with it a sense of urgency that can accelerate decision-making and innovation. During the pandemic, organizations were forced to quickly adapt to new ways of working. Remote work, which was once considered a perk or a novelty, became the norm almost overnight. This sudden shift presented numerous challenges, but it also offered a unique opportunity for organizations to experiment with new ways of working and to learn from these experiments in real-time (Russo et al., 2023a).

The pandemic has underlined the need for agility and flexibility in organizations. Traditional, hierarchical organizational structures have shown their limitations in the face of rapid and large-scale disruptions. Agile structures, which are characterized by decentralized decision-making and a high degree of autonomy at the team level, have proven more effective at adapting to the new reality (Cucolaş & Russo, 2023). This shift towards agility represents a significant learning from the pandemic that is likely to shape the future of work even after the crisis subsides.

At the same time, the pandemic has highlighted the importance of employee wellbeing and mental health. Organizations have learned that supporting their employees' mental health is not just a moral imperative, but it is also essential for maintaining productivity during stressful times (Stander & Zyl, 2023). This recognition is prompting organizations to invest more in mental health resources and to foster a culture that prioritizes wellbeing (Russo et al., 2021).

While the COVID-19 pandemic has presented significant challenges, it has also offered valuable learnings. The experience has underscored the importance of resilience, agility, digital readiness, and employee wellbeing. As we move forward, it is critical that we continue to learn from these disruptions and incorporate these learnings into our strategies and practices. By doing so, we can turn the challenges of today into the opportunities of tomorrow, bouncing forward to capture sustainable, inclusive growth.

## The Integrated Job Demands-Resources and Self-Determination Model (IJARS)

The past few years have been a period of unprecedented change and challenges, particularly for the software development industry. As we navigate the wake of the COVID-19 pandemic, understanding and nurturing the wellbeing and resilience of software developers has become a paramount concern. This leads us to propose the Integrated Job Demands-Resources and Self-Determination Model (IJARS), a comprehensive framework that sheds light on the wellbeing and productivity of software engineers during these transformative times (Russo et al., 2023a), summarized in Figure 1.

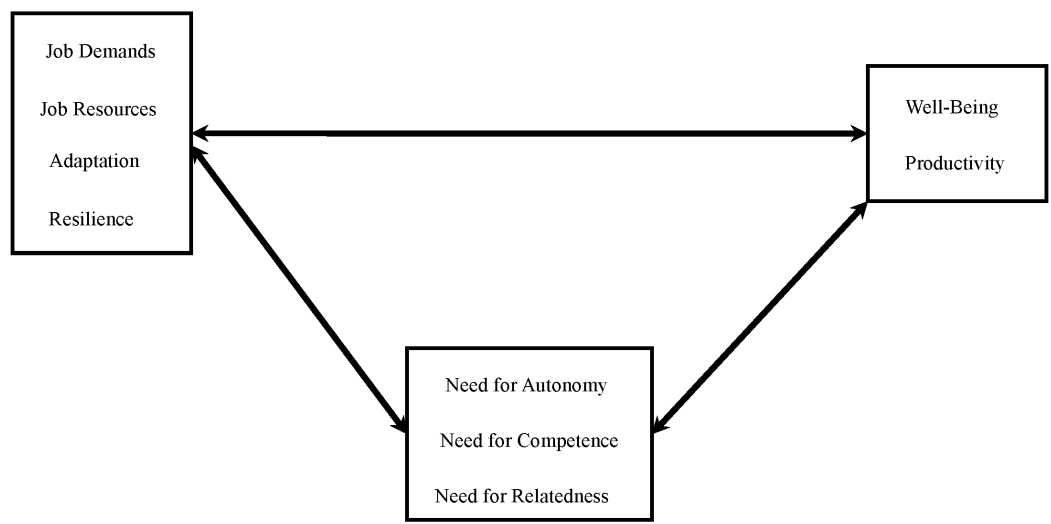

*Figure 1 The Integrated Job Demands-Resources and Self-Determination Model from Russo et al., 2023a*

The Integrated Job Demands-Resources and Self-Determination Model combines elements from both the Job Demands-Resources (JD-R) Model by Demerouti et al. (2001) and Self-Determination Theory (SDT) by Deci et al. (2000), two influential concepts in occupational psychology. Job demands refer to the challenging aspects of a job, such as workload or role ambiguity, which could undermine wellbeing and productivity. On the other hand, job resources, like management support or skill development opportunities, can enhance these aspects, promoting a more positive work environment.

The Job Demands-Resources Model, a critical part of our framework, acknowledges that work environments are characterized by job demands and job resources. High job demands can lead to stress and burnout, while ample job resources foster work engagement and positive outcomes. It's a delicate balance that requires thoughtful consideration from managers and organizations.

Self-Determination Theory adds another layer to our understanding by focusing on three basic psychological needs: autonomy, competence, and relatedness. These needs are essential for motivation, performance, and wellbeing in various domains, including the workplace. When employees feel a sense of choice in their actions, feel effective in their abilities, and feel connected to others, they are more likely to thrive.

The COVID-19 pandemic has brought about a significant shift in work environments, introducing new challenges for software engineers. To cope with these changes, adaptation and resilience have become vital. Adaptation refers to adjusting to new or altered circumstances, while resilience refers to the capacity to bounce back from adversity. Both these qualities have been essential for software engineers to maintain their wellbeing and productivity during the pandemic.

What sets the Integrated Job Demands-Resources and Self-Determination Model apart is its nuanced understanding of how job demands, resources, and basic psychological needs interact, along with the roles of adaptation and resilience. It presents a holistic view, recognizing that software engineers' wellbeing and productivity are influenced by a complex interplay of these factors.

The Integrated Job Demands-Resources and Self-Determination Model paints a vivid picture of the intricate connections between these components and how they collectively contribute to the wellbeing and productivity of software engineers. It underscores the need for organizations to identify and address critical job demands, enhance job resources, and foster an environment that supports autonomy, competence, and relatedness.

In conclusion, the Integrated Job Demands-Resources and Self-Determination Model offers a multifaceted perspective on the wellbeing and productivity of software engineers in the wake of the pandemic. By synthesizing concepts from occupational psychology with the real-world experiences of software developers, it provides a robust framework. Its insights serve as a vital guide for organizations, researchers, and practitioners, helping to nurture resilience and transform adversity into opportunity in the ever-evolving world of software development.

**Conclusion: A New Blueprint for the Future**

The COVID-19 pandemic has underscored the critical roles of wellbeing and resilience in software development. Our research emphasizes a holistic approach, encapsulated in the Integrated Job Demands-Resources and Self-Determination Model (IJARS), which considers job demands, resources, psychological needs, adaptation, and resilience.

The pandemic's challenges have revealed the effectiveness of Agile values like Scrum in remote work, highlighting their ability to enhance collaboration and communication. Our findings stress the importance of a learning mindset, openness to new ways of working, and the opportunity to reshape our industry, prioritizing wellbeing and resilience. These lessons create a promising future where the software development field can act as a blueprint for other sectors in our increasingly digital world.

In conclusion, we can forge a more robust and resilient software development industry by prioritizing wellbeing, embracing Agile values, and fostering a learning mindset.

## Acknowledgments

This paper summarizes the lessons learned from the PanTra — Pandemic Transformation project, generously funded by Carlsberg Foundation under grant agreement number CF20-0322. I would like to extend my deepest gratitude to my co-authors, listed alphabetically, who made this research possible: Seraphina Altnickel, Niels van Berkel, Adrian-Alexandru Cucolaș, Emily Laue Christensen, Paul H. P. Hanel, Darja Šmite, and Paolo Tell. ChatGPT-4 has been used to ensure linguistic accuracy and enhance the readability of this article.

## Actionable Insights

1. **Integrate Wellbeing and Resilience Through the IJARS Model**: The manuscript introduces the Integrated Job Demands-Resources and Self-Determination Model (IJARS), emphasizing the interplay between job demands, resources, adaptation, and resilience. Software practitioners should utilize this model to understand and foster mental wellbeing, implementing measures such as mental health resources, flexible work arrangements, and a supportive culture.
2. **Adopt Agile values**: Agile methodologies, especially Scrum, have proven effective in remote work scenarios. Implementing these frameworks and their values can enhance collaboration, communication, and overall team performance. This supports the IJARS model by aligning job demands and resources with individual needs.
3. **Embrace Hybrid Work Arrangements**: Hybrid work models, blending in-office and remote work, can contribute to the wellbeing of software developers. The manuscript emphasizes the importance of these models in promoting work-life balance and aligning with the IJARS framework. Success in hybrid work requires careful planning, transparent communication, and robust support systems.

# Bio

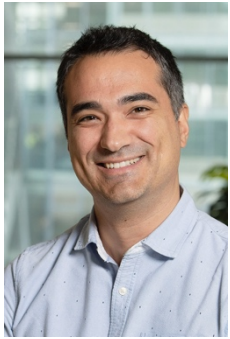

DANIEL RUSSO is an associate professor of computer science at Aalborg University in Copenhagen, Denmark. His research interests include the broad area of empirical software engineering, studying the effects of the COVID-19 pandemic on software professionals, Agile software development, Generative AI adoption, and research methodology. Daniel Russo received a Ph.D. in computer science and engineering from the University of Bologna. Contact him at www.danielrusso.org or daniel.russo@cs.aau.dk.